\documentstyle[floats,multicol,aps,epsfig,prl,subfigure]{revtex}

\newlength{\bxwidth}\bxwidth=3.2 truein
\newcommand\om { \omega}

\newlength{\jight}
\newlength{\fight}
\newcommand{\fg}[4]
{\begin{figure}[h]\epsfxsize=#1
\centerline{\epsfbox{#2}}\vskip 0.1truein
\caption{{#3}}\label{#4}\end{figure}}
\newcommand\ltdash{\raise-1.8pt\hbox{$\scriptscriptstyle |$}}
\newcommand \beq  {\begin{equation}}
\newcommand \eeq  {\end{equation}}
\newcommand \bea {\begin{eqnarray} }
\newcommand \eea {\end{eqnarray}}

\newcommand \dg{\dagger}

\newcommand\bk{{\bf k}}

\newcommand\bq{{\bf q}}

\newcommand\bv{{\bf v}}
\newcommand\tx{\textstyle}

\newcommand\etal{{\it et al.}}

\begin{document}
\draft
\twocolumn[\hsize\textwidth\columnwidth\hsize\csname @twocolumnfalse\endcsname
\title{
Fractionalization and Fermi surface volume in heavy fermion
compounds: the case of YbRh$_2$ Si$_2$.}
\author{  Catherine P\'epin  }
\address{ SPhT, L'Orme des Merisiers, CEA-Saclay, 91191 Gif-sur-Yvette, France
}

\maketitle
\date{\today}
\maketitle
\begin{abstract}
We establish an effective theory for heavy fermion compounds close
to a zero temperature Anti-Ferromagnetic (AF) transition. Coming
from the heavy Fermi liquid phase across to the AF phase, the
heavy electron fractionalizes into a light electron, a bosonic
spinon and a {\it new} excitation: a spinless fermionic field.
Assuming this field acquires dynamics and dispersion when one
integrates out the high energy degrees of freedom, we give a
scenario for the volume of its Fermi surface through the phase
diagram. We apply our theory to the special case of
YbRh$_2$(Si$_{1-x}$ Ge$_x$)$_2$ where we recover, within
experimental resolution, several low temperature exponents for
transport and thermodynamics.

\end{abstract}
\vskip 0.2 truein 
\newpage
\vskip2pc]


In some heavy fermion compounds, pronounced deviations from the
conventional Landau Fermi liquid behavior have been reported when
the compounds are tuned through an antiferromagetic (AF) quantum
critical point (QCP) \cite{review}. Generically, the specific heat
coefficient is seen to diverge at the QCP, showing at least
logarithmic increase \cite{lohneysen,steglich,aeppli,movshovich}
as the temperature is decreased. The resistivity is quasi-linear
in temperature \cite{grosche,gegenwart,julian,paglione}. NMR and
$\mu$-SR studies, as well as neutron scattering measurements for
one compound \cite{aeppli} show that the spin susceptibility
acquires some anomalous exponent \cite{grosche,gegenwart}. Here we
focus on the special case of YbRh$_2$(Si$_{1-x}$ Ge$_x$)$_2$ doped
with Ge to reach a QCP at $ x=0.05 $. In this compound, linear
resistivity is observed, at the QCP, over three decades of
temperature, from 10 K to 10 mK \cite{jeroenphd}. At the same
time, the specific heat coefficient shows an upturn below 300 mK
from its original logarithmic increase, $C/T \sim T^{- \alpha}$
with $ \alpha \simeq 0.33 $ \cite{gegenwart,custers}. The entropy
associated with  the upturn is of the order of 5 $\%$ of the total
spin entropy of the Yb atom. Curiously, the upturn in specific
heat is un-correlated with transport, since one doesn't see any
kink at 300 mK, or any deviation from the linear slope in the
resistivity. An NMR study on the pure compound \cite{ishida} shows
that the relaxation time on Si $ 1 / T_1 T \sim T^{-1/2} $ when
the system is driven close to the QCP with applied magnetic field,
suggesting that $\sum_q \chi^{\prime \prime} (q, \omega)/ \omega
\sim T^{-0.5}$. The Gr{\"u}neysen parameter, ratio of the thermal
expansion coefficient and the specific heat is shown to diverge
with an unusual exponent, i.e. $\Gamma = \beta/ C \sim T^{-0.7}$,
compared to the case of CeNi$_2$Ge$_2$, well understood within a
Spin Density Wave(SDW) scenario \cite{kuchler}. Latest
measurements of the Hall constant \cite{silke} suggest that the
Fermi surface volume is increasing, when one goes from the AF to
the field induced heavy Fermi liquid phase. A recent study of the
heavy Fermi liquid phase shows a generic scaling in $B/T$ in the
transport and specific heat \cite{custers}. One also observes that
the Kadowaki-Woods ratio -- $ K= A/\gamma^2 $ where $A$ is the
$T^2$ coefficient of the resistivity and $\gamma =C/T|_{T
\rightarrow 0}$ is the specific heat coefficient-- increases in
approaching the QCP. Likewise the Wilson ration -- $W= \chi/\gamma
|_{T \rightarrow 0}$ where $\chi$ is the bulk susceptibility--
shows a dramatic increase. In contrast, the ratio of $A/ \chi^2$
stays constant over the whole magnetic field range. We deduce from
the fact that the $A$ coefficient doesn't feel the upturn in
specific heat, that the bulk susceptibility $\chi$ doesn't feel it
either --their ratio being constant \cite{constantAchi}. In other
words, the excitations responsible for the upturn seen in $C/T$
don't couple to the bulk susceptibility.

In this paper we introduce an effective theory for heavy fermion
compounds, which, in the particular case of YbRh$_2$Si$_2$,
accounts for the whole set of experimental evidence quoted above.


Our starting point is the Kondo-Heisenberg lattice model, where we
use a Schwinger representation for the spin of the impurities,
which we refer to as the bosonic Kondo-Heisenberg (BKH) model. The
BKH lattice Hamiltonian \bea \label{BKH}
H  =   H_c  + & H_K & + H_H \  ,  \nonumber \\
\mbox{where} \ H_c & = & \sum_{k \sigma} \varepsilon_k
f^\dagger_{k \sigma} f_{k
\sigma} \ ,  \nonumber \\
H_K & =&  J_K \sum_{i \sigma \sigma^\prime} b^\dagger_{i \sigma}
b_{i \sigma^\prime} f^\dagger_{i \sigma^\prime} f_{ i \sigma} \ , \nonumber  \\
H_ H & = &  J_H \sum_{(i,j) \sigma \sigma^\prime} b^\dagger_{i
\sigma} b_{i \sigma^\prime} b^\dagger_{j \sigma^\prime} b_{j
\sigma} \eea describes the conduction band ($H_c$), the Kondo
coupling between local moments and the conduction electrons at
site $i$ ($H_K$), and the super-exchange between neighboring spins
($H_H$). When we formulate the BKH model as a functional integral,
we can decouple the fields as follows \bea \label{eq2} H_K &
\rightarrow & H^\prime_K = \sum_{i \sigma} \left [ b^\dagger_{i
\sigma} \chi_i^\dagger f_{i \sigma} + h.c. \right ] -
\frac{\chi^\dagger_i \chi_i}{J_K} \\
H_H & \rightarrow & H^\prime_H =\sum_{(i,j) \sigma} \left [
|\Delta_{ij}| e^{i \frac{\pi}{a} {\bf (r_i- r_j )}} b^\dagger_{i
\sigma} b_{j -\sigma}^\dagger + h.c. \right] - \frac{|
\Delta_{ij}|^2}{J_H} \ , \nonumber \eea where the on-site bond
variable in the first term is a Grassman field which doesn't carry
a spin and the bond variable in the second term has been chosen
following an SP(2) decomposition of the
interaction~\cite{sachdev-spn}.

We believe the fermion field $\chi$  is a {\it good} decoupling of
the Kondo interaction in the sense that it describes the set of
elementary excitations close to the QCP \cite{cathchifermion}.
Through coupling to the spinons ( $b_{k \sigma}$) and the
itinerant electrons ($f_{k \sigma}$) the $\chi$-fermion Kondo bond
states acquire some dynamics, damping and dispersion. Within this
set of variables, the formation of the Kondo heavy quasiparticle
is described with the formation of a ``band'' for the
$\chi$-fermion bound state. To see how this happens it is
instructive to step back to the impurity case.

Magnetism in the bosonic Kondo impurity model has been
investigated in the past \cite{oldpiers,olivier}, but here we
focus on the fate of the $\chi$-fermion chemical potential
\cite{cathkondo}. From (\ref{eq2}), one notices that, for the
antiferromagnetic Kondo effect ($J_K >0$), the energy level is
negative, thus full, and the chemical potential $\mu= - 1/ J_K$
flows to zero at strong coupling. The $\chi$-fermion becomes
massless. This singularity accounts for the increase of the Fermi
surface volume by one. Oppositely, in the ferromagnetic case ($J_K
<0$), the energy level is positive, thus empty, and the chemical
potential $ \mu= -1 / J_K$ flows to infinity at low energies,
leaving the Fermi surface volume unchanged.

\begin{figure}
\epsfig{figure=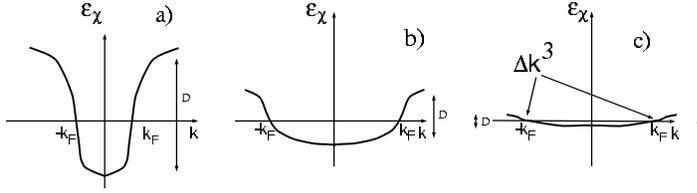,height=3.2 cm} \caption{ Evolution of
the Fermi surface volume from the AF phase to the heavy Fermi
liquid phase. $D$ is the bandwidth.
 }  \label{dispers2}
\end{figure}

In the Kondo lattice, the $\chi$-fermion acquires some dynamics,
damping and dispersion. The competition between AF fluctuations
and the Kondo screening affects the formation of the $\chi$-band.
Moreover, one can imagine that the lattice accommodates partial
screening of the impurity spin through the $\chi$-dispersion: the
``unscreened'' part of the spin --by analogy with ferromagnetic
Kondo-- has positive energies, while the screened part of the spin
has negative ones. Deep inside the AF phase (diagram a in Fig.
\ref{dispers2} ), the spin of the impurities are partially
screened and the $\chi$-fermion bandwidth is large -- partially
filled band, with a rather ``small'' Fermi surface.
 As the strength of AF fluctuations is
decreased, the number of empty energy levels decrease compared to
full energy levels, the $\chi$-band becomes {\it flatter} and the
Fermi surface is bigger (diagram b in Fig. \ref{dispers2}). Deep
inside the heavy Fermi liquid phase, the Kondo effect is dominant,
and one expects a situation similar to the one impurity case, with
a totally flat $\chi$-band and a large Fermi surface; in this
phase the $\chi$-fermion is no longer a good elementary excitation
and, $\chi$ being a flat massless mode, the theory is ill defined.
Just at the brink of the Fermi liquid phase, the $\chi$-band is
``almost'' flat (diagram c in Fig. \ref{dispers2}). One can thus
expect that the velocity vanishes for  part of the Fermi surface,
producing a singularity.

\begin{figure}
\epsfig{figure=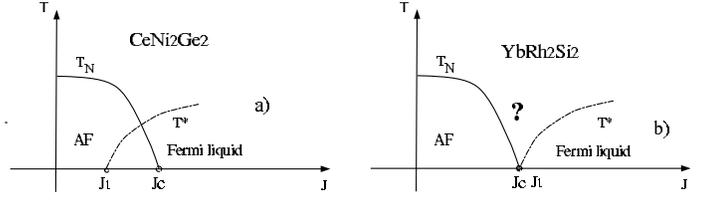,height=3.3 cm} \caption{Two possible
phase diagrams in heavy fermion compounds.
 }  \label{phases}
\end{figure}

To summarize, our scenario predicts a new excitation close to the
QCP, which is fermionic in nature and doesn't carry a spin. As one
integrates out the high energy degree of fredom, this new fermion
forms a ``band'': its dispersion has positive and negative energy
levels. The band structure at the QCP is not universal, it depends
on the relative strengths of the Kondo effect and AF interactions
(Fig. \ref{phases}).
If the AF interaction is small compared to the Kondo interaction,
the QCP lies inside the heavy Fermi liquid phase, which is
expected to be the case for CeNi$_2$ Ge$_2$ (diagram a  in Fig.
\ref{phases}). In the opposite limit, when the AF fluctuations are
strong compare to the Kondo effect, the Fermi liquid phase can lie
{\it outside} the AF phase, and one can expect singularities at
the QCP (diagram b in Fig. \ref{phases}). This is the case for the
YbRh$_2$Si$_2$ compound. The evolution of the $\chi$-band
structure accounts for the variation of the total Fermi surface
volume, with a smooth cross-over from a rather small Fermi surface
inside the AF phase to a big Fermi surface inside the Fermi liquid
phase.

\begin{figure}
\epsfig{figure=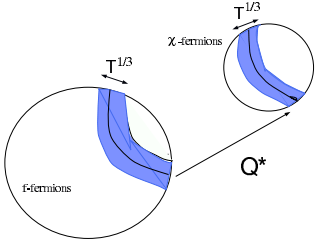,height=3.5 cm} \caption{ The
$f$-fermion and $\chi$-fermion Fermi surfaces. The hot lines are
points related by the ordering wave vector ${\bf Q}^*$; their
width is proportional to $T^{1/3}$.
 }  \label{figure3}
\end{figure}

We now turn to the specific case of YbRh$_2$(Si$_{1-x}$
Ge$_x$)$_2$ doped to criticality for $x=0.05$. We assume that, at
the QCP, the $\chi$-band is well formed, but the velocity vanishes
at the hot lines. The effective Lagrangian comprises four terms: a
conduction electron term $S_f$, a boson term $S_b$ which describes
the critical spinons, the $\chi$-fermion term $S_\chi$, and the
interaction $S_{int}$ between those modes. Both the $f$ and
$\chi$- fermion's Fermi surface have hot lines related by the
ordering wave vector $Q^*$ (Fig.\ref{figure3}).
     \bea
\label{eqn1} S &  = & S_f + S_\chi +
S_b + S_{int} \\
     S_f & = & \int \frac{d \om \ d^3 k}{(2 \pi)^4} \sum_\sigma \ f^{\dg}_{\bk \sigma} \
\big( i \om -\bv_F \cdot {\bf k}
  \big) \ f_{\bk \sigma} \nonumber \\
  S_\chi & = & \int \frac{d \om \ d^3 k}{(2 \pi)^4} \ \chi^{\dg}_{\bk} \
\big( i \Sigma_\chi - \bv_\chi \cdot \bk + {\bf a} \cdot \bk^3
  \big) \ \chi_\bk \nonumber \\
     S_b & = & \int \frac{d \nu \ d^3 q}{ (2 \pi)^4} \sum_\sigma \;
     \ b^\dg_{\bq \sigma} \ \big(
 i \nu - \lambda q \big) b_{\bq \sigma}
\nonumber \\
    S_{int} & = & g \int d \om_1 \ d \om_2 \  d^3 k_1 \ d^3 k_2 \ d^3 q \
\delta_{\bq + \bk_1 -\bk_2} \nonumber \\
~ & ~ & \qquad \sum_\sigma ( b^\dg_{\bq \sigma} \psi^{\dg}_{\bk_2}
f_{\bk_1 \sigma} + h.c.)
 \nonumber
 \ , \eea where the $f$-fermion as well as the $\chi$-fermion
 dispersions have been linearized around the Fermi surface, the
 Fermi velocity
 $\bv_\chi$ vanishes at the hot lines, leading to a singular in $\bk^3$ dispersion.
 We take $\Sigma_\chi= \omega \log(
 |\omega |)$. It is assumed the width of the $\chi$-fermion hot lines varies
 with temperature like
 $\Delta q = (T/a)^{1/3}$.
 The spinon dispersion is linearized close to the AF QCP , after
 diagonalizing $H_H$ in (\ref{BKH}) in a mean field
 approximation: $\omega_q = \sqrt{\lambda^2 - \Delta_q^2}$ with
 $\lambda$ the Lagrange parameter implementing the constraint on
 the SP(N) representation of the spin, and $\Delta_q= \Delta (\sin
 q_x + \sin q_y + \sin q_z)$. At the QCP, $\lambda= \Delta$ and we
 have linearized $\omega_q $ around $\bq= (\pi/2,\pi/2,\pi/2)$.
We first evaluate the dynamical vertex $g(\epsilon, \Omega)$
around the hot lines, where $\epsilon$ and $\Omega$ are the
incoming fermion and boson frequencies respectively and find \beq
g(\omega, \nu) \sim Max[ \omega, \nu]^{5/6} \ . \eeq The effective
theory (\ref{eqn1}) is thus stable with respect to the interaction
$g$. It lies above its upper critical dimension. Critical
exponents can be evaluated by computing self-consistently the
self-energies at one loop (Fig. \ref{diag}), the smallness of the
vertex ensuring the validity of this treatment. Solving the set of
three Dyson equations for the three fields $b$, $f$ and $\chi$
leads to momentum independent self-energies \beq \label{self}
\begin{array}{l}
\Sigma_b(\nu) \simeq \frac{g^2}{v_F}  \left [ \frac{|\nu|}{a} \log
\frac{\nu}{v_F} \right ]^{1/3} \ ; \\ \Sigma_f(\omega) \simeq i
\mbox{sgn} (\omega) \frac{g^2 }{\lambda} \left [\frac{|\omega|}{a}
\log \frac{\omega}{ v_F} \right ]^{2/3}  ;
\end{array} \eeq $ \Sigma_\chi (\omega) \sim i | \omega|^{4/3} sgn
(\omega)$ being irrelevant compared to the original damping.
\fg{\bxwidth}{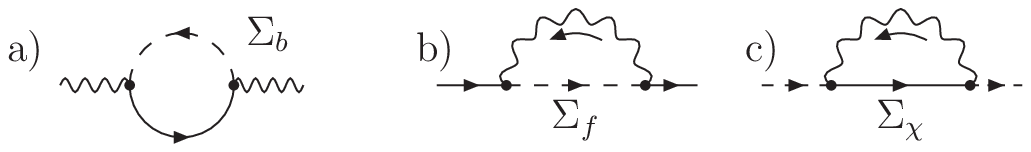}{Feynman diagrams associated with a) the
spinon
 self-energy b) the $f$-fermion self energy c) the $\chi$-fermion
 self-energy.
  Solid, dashed and wavy lines are respectively $f$-fermion, $\chi$-fermion  and boson
propagators.}{diag}

In order to discuss the thermodynamics of our model, we rely on a
scaling argument. The free energy is
 comprised of three terms
  $ F = F_{f} + F_{\chi} +
F_{bos} $ , where $F_f$ is the free energy of the heavy
$f$-fermions, $F_\chi$ is the contribution from the $\chi$
fermions and $F_{bos}$ from the boson modes. Any of these
individual contributions is extensive and  proportional to the
phase space volume $F \sim T \prod_i \Delta q_i $, where $\Delta
q_i$ is the momentum width in the $i$-direction. For the
$\chi$-fermions, only two directions, perpendicular to the hot
lines, reduce the phase space with $\Delta q_{1,2} \sim
(T/a)^{1/3}$ leading to $ F_\chi \sim T (T/a)^{2/3 } $. The boson
modes scale in all
 directions, leading to $F_{bos} \sim T (\Sigma_b(T)/ \lambda)^3
 \sim T^2$. The $f$- fermions scale in two
 directions, perpendicular to the hot lines, with $\Delta q_1 \sim
 (T/a)^{1/3}$ and $\Delta q_2 \sim \Sigma_f(T)/ v_F$, leading to
 $F_{f , hot} \sim T^2$. Notice that the scaling exponents of the
 hot fermion's and the boson's entropy are the same, which is a
 property of spin-fermion models above their upper critical
 dimension \cite{chubukov}. At low temperatures, $F_\chi$ is the dominant
 contribution, leading to a specific heat coefficient which
 captures the upturn observed in the experiments:
 \beq \gamma = C/T \sim a^{2/3} T^{-1/3} \ . \eeq

\begin{figure}
\epsfig{figure=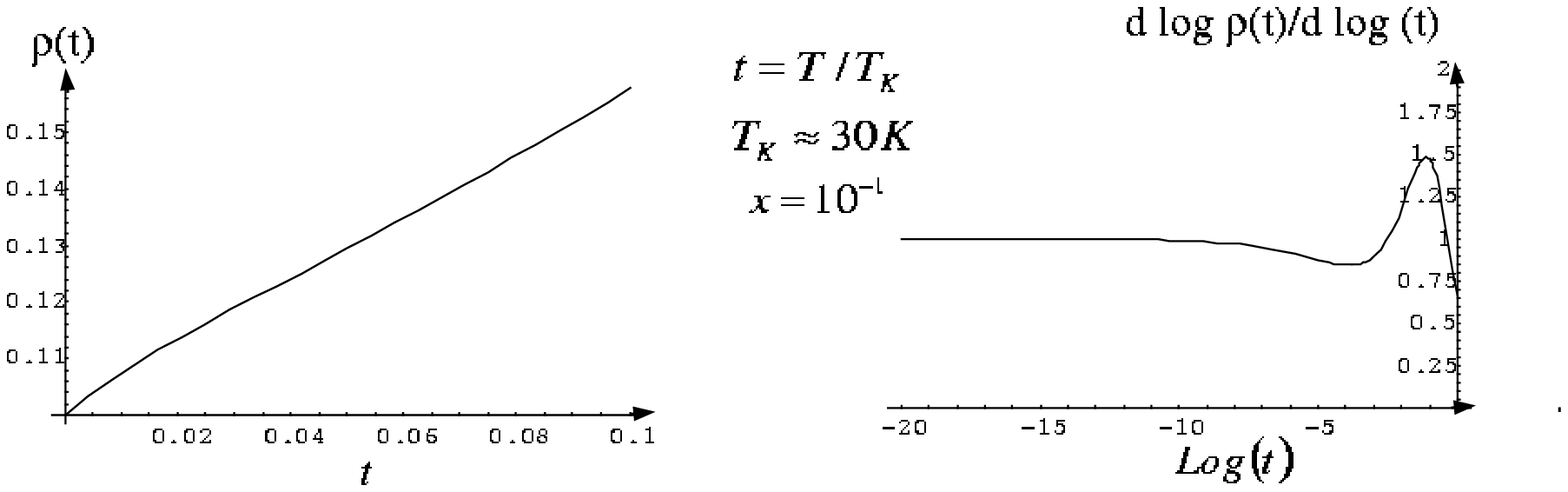,width=9 cm} \caption{Linear in $T$
resistivity and variation of the temperature exponent as function
of $t=T/ T_K$. $x$ is the inverse Residual Resistivity Ratio.
 }  \label{transport}
\end{figure}

 We now turn to the behavior of the conductivity at
 criticality. The $f$-fermions are the sole
contributors to electric transport because the velocity of the
$\chi$-fermion vanishes on the hot lines. Assuming that the
scattering processes from magnetic spinons dominate the transport,
one distinguishes the contributions from the cold and hot regions
$ \sigma( T) = \sigma_{cold} (T) + \sigma_{hot}(T) $. Following
\cite{rosch}, the conductivity is the sum of inverse scattering
rates associated with the hot and cold regions, multiplied by the
phase space allowed to these regions
 \[
\begin{array}{ll}
\sigma_{cold} \sim \frac{ \tx 1 - t^{1/3} }
 {\tx  x + t^2 }  \ ; &
\sigma_{hot} \sim \frac{ \tx t^{1/3} } {\tx x + t^{2/3}} \ ,
\end{array} \] where $x$ is a dimensionless
parameter which characterizes the residual resistivity-- for
example, the inverse of the Residual Resistivity Ratio (RRR),
being the ratio of the resistivity  at 100 K to the resistivity at
10 K. $t$ is a dimensionless parameter characteristic of the
temperature --one can take for example $t= T/ T_K$, where $T_K$ is
the one impurity Kondo temperature of the compound. $t^{1/3}$ is
the width of the hot lines, $t^2$ is the  inverse scattering rate
of the cold conduction electrons, and $t^{2/3}$ is the inverse
scattering rate of the conduction electrons -- up to logarithms--
from the paramagnetic spinons around the hot lines. One sees three
regimes in the conductivity: \beq
\begin{array}{ccc}
t < x^{3/2} \ ; & x^{3/2} < t < x^{1/2} \ ; & x^{1/2} < t \\
\rho= x +t ; &  \mbox{cross-over} \; \; \; \; \; \; \; \; \; ; &
\Delta \rho \sim t^2 \ .
\end{array} \eeq With $x \simeq 0.1$ and $T_K \simeq
30 K$ for YbRh$_2$Si$_2$, one sees that the resistivity is linear
in $T$, below $T= 3 K$ down to the lowest temperature
(Fig.\ref{transport}), while the $T^2$ behavior would appear
roughly above $ T=10 K $, corresponding to the experimental
observation.
To conclude the study of thermal observables, we focus on the
Gr{\" u}neysen parameter, defined as $\Gamma = \alpha/ C_p$ with
$\alpha \sim \frac{\partial F}{ T \partial p}$ and $ C_p \sim
\frac {\partial F}{\partial T}$. $p$ is the hydrostatic pressure.
Calling $r=(p-p_c)/p_c$ the departure from the QCP, and noticing,
from a small displacement of the $\chi$-fermion hot line that $r
\sim q_{typ}^2 \sim T^{2/3}$, one finds $\Gamma \sim T^{-2/3}$
close to the QCP. This exponent is in good agreement with the
experiments \cite{kuchler}.

 Several theoretical descriptions have been advanced to account for the
 striking experimental results around the QCP of heavy fermions.
 Anisotropic (2D) spin density wave scenario\cite{2dsdw,si} can explain the linear in $T$
 resistivity, as well as the logarithmic dependence of the specific
 heat coefficient. A theory with a ``local'' mode, as well as 2D spin fluctuations
  at criticality, has been advocated \cite{si}, which also captures
   the anomalous exponent of the spin susceptibility. None of those
  theories can account for the upturn in the specific heat
  coefficient.
  Our approach has some analogies with the idea of
  ``fractionalization'' \cite{senthil} and deconfinement
  \cite{review}, in the  sense that a new excitation appears at the
  QCP, which is fermionic in nature, and which, at high energies,
  is the gauge field which characterizes the Kondo bound state.
  When the new excitation is fermionic in nature, one might expect
  the emergence of a new ``phase'' where the excitation is stabilized
  \cite{cathchifermion,senthil-vish}. Some analogies exist with the
  ``two-fluid model'' advanced recently to describe CeCoIn$_5$
  \cite{pines}: in our model also, the impurity
  spins are partially screened, the difference being that at very
  low energies, the unscreened part of the spin, instead of remaining intact,
   ``fractionalizes''
  into a spinon, a spinless fermion, and a ``light'' electron.

  To summarize, starting from the Kondo-Heisenberg model at high
  energies, we have introduced a new effective theory for the QCP
  of YbRh$_2$(Si$_{0.95}$ Ge$_{0.05}$)$_2$. This effective theory
  shows the emergence of a new excitation at low energies, fermionic
  and spinless
  in nature, characterizing the Kondo bound state. This new fermion
  --the $\chi$-fermion-- forms a ``band''  when high energy degrees of
  freedom are integrated out.
  The total Fermi surface volume evolves from a small Fermi surface
  to a big Fermi
  surface as the $\chi$-fermion band fills up.
  Assuming a vanishing velocity in the $\chi$-fermion
  band structure
  close to the QCP, one can reproduce many striking experimental
  observations of YbRh$_2$(Si$_{0.95}$ Ge$_{0.05}$)$_2$. One
  captures the upturn in the specific heat coefficient, with the
  right exponent with temperature, as well as the linear
  resistivity at very low temperatures. The upturn in the specific heat
  coefficient doesn't couple to
  transport, and couples only indirectly --via the spinons and $f$-fermions--
   to the bulk spin susceptibility. This explains the observation that the ratio
    $A/ \chi^2$ is constant in the heavy Fermi liquid phase \cite{constantAchi}.
   One also reproduces the variation of
  the Gr{\" u}neysen parameter with the anomalous temperature
  exponent.

  I would like to thank  A.V. Chubukov, P. Coleman, J. Custers, P. Gegenwart,
  K. Le Hur, M. Norman, I. Paul, O. Parcollet,
S. Pashen, J. Rech, and F. Steglich for many discussions related
to this work. A special thanks to the Max Planck Institut f{\" u}r
Chemische Physik fester Stoffe in Dresden, for their hospitality
and where part of this work was completed.

\end{document}